# The Role of Radiographic Knee Alignment in Total Knee Replacement Outcomes and Opportunities for Artificial Intelligence-Driven Assessment

Zhisen Hu, Dominic Cullen, David S. Johnson, Aleksei Tiulpin, Timothy F. Cootes, and Claudia Lindner

***Abstract*—Knee osteoarthritis (OA) is one of the most widespread and burdensome health problems [1-4]. Total knee replacement (TKR) may be offered as treatment for end-stage knee OA. Nevertheless, TKR is an invasive procedure involving prosthesis implantation at the knee joint, and around 10% of patients are dissatisfied following TKR [5,6]. Dissatisfaction is often assessed through patient-reported outcome measures (PROMs) [7], which are usually completed by patients and assessed by health professionals to evaluate the condition of TKR patients. In clinical practice, predicting poor TKR outcomes in advance could help optimise patient selection and improve management strategies.**

**Radiographic knee alignment is an important biomarker for predicting TKR outcomes and long-term joint health. Abnormalities such as femoral or tibial deformities can directly influence surgical planning, implant selection, and postoperative recovery [8,9]. Traditional alignment measurement is manual, time-consuming, and requires long-leg radiographs, which are not always undertaken in clinical practice. Instead, standard anteroposterior (AP) knee radiographs are often the main imaging modality. Automated methods for alignment assessment in standard knee radiographs are potentially clinically valuable for improving efficiency in the knee OA treatment pathway.**

## I. Introduction

KNEE osteoarthritis (OA) is one of the most widespread and burdensome health problems [1-4]. Total knee replacement (TKR) may be offered as treatment for end-stage knee OA. Nevertheless, TKR is an invasive procedure involving prosthesis implantation at the knee joint, and around 10% of patients are dissatisfied following TKR [5,6]. Dissatisfaction is often assessed through patient-reported outcome measures (PROMs) [7], which are usually completed by patients and assessed by health professionals to evaluate the condition of TKR patients. In clinical practice, predicting poor TKR outcomes in advance could help optimise patient selection and improve management strategies.

Radiographic knee alignment is an important biomarker for predicting TKR outcomes and long-term joint health. Abnormalities such as femoral or tibial deformities can directly influence surgical planning, implant selection, and postoperative recovery [8,9]. Traditional alignment measurement is manual, time-consuming, and requires long-leg radiographs, which are not always undertaken in clinical practice. Instead, standard anteroposterior (AP) knee radiographs are often the main imaging modality. Automated methods for alignment assessment in standard knee radiographs are potentially clinically valuable for improving efficiency in the knee OA treatment pathway.

Machine learning and deep learning have been widely applied in medical image analysis including for knee alignment assessment [10-13]. However, many current review articles in knee image analysis focus on knee OA diagnosis [14-18] or MRI-based knee segmentation [19-22], not considering the assessment of knee alignment in knee radiographs. To address this gap, this article provides a comprehensive review of the current literature on radiographic knee alignment factors influencing TKR outcomes and on AI-based techniques for automated alignment assessment.

**Contributions**: i) We explore possible radiographic knee alignment biomarkers for TKR outcomes and relevant techniques for their assessment. ii) We investigate and summarise the current literature on AI-based techniques used to assess radiographic knee alignment for TKR outcomes. iii) We discuss available commercial products for automated knee alignment measurement. iv) We identify gaps and possible future directions for generating automated knee alignment biomarkers for TKR outcomes and predictions of TKR outcomes.

ZH is funded by European Laboratory for Learning and Intelligent Systems (ELLIS) Unit Manchester. CL is funded by a Sir Henry Dale Fellowship jointly funded by the Wellcome Trust and the Royal Society (223267/Z/21/Z), DC was supported by the same award.

*Z. Hu is with The University of Manchester, United Kingdom, and also with the University of Oulu, Finland. (correspondence e-mail: zhisen.hu@postgrad.manchester.ac.uk).

D. Cullen, T. F. Cootes, and C. Lindner are with The University of Manchester, United Kingdom.

D. S. Johnson is with Stockport NHS Foudation Trust, Stepping Hill Hospital, Stockport, United Kingdom.

A. Tiulpin is with the University of Oulu, Finland.

TABLE I
SCORING PROTOCOLS FOR PATIENT-REPORTED OUTCOME MEASURES REGARDING KNEE REPLACEMENT SURGERY

| Scoring Protocols | Assessment | Grading Scale |
|---|---|---|
| KSS | Knee pain, RoM, stability, walking function, stair climbing function | Clinical rating system (0-100): knee pain 0-50, RoM: 0-25 points, stability: 0-25.<br>Functional scoring system (0-100): walking: 0-50, stair climbing: 0-50.<br>The higher the better. |
| OKS | 12 questions regarding knee health, pain severity, stiffness, daily activities | Excellent (41-48): minimal symptoms as well as excellent knee function.<br>Good (34-40): some symptoms, but good knee function.<br>Fair (27-33): moderate symptoms, and knee function.<br>Poor (0-26): significant symptoms, and poor knee function. |
| KOOS | Impact of knee injury and OA on an individual's quality of life | Pain: 0-100.<br>Symptoms: 0-100.<br>Activities of daily living: 0-100.<br>Sport and recreation function: 0-100.<br>Knee-related quality of life: 0-100.<br>The higher the better. |
| WOMAC | 24 questions regarding function, pain, and stiffness | Pain subscale: 5 questions, 0-20.<br>Stiffness: 2 questions, 0-8.<br>Physical function: 17 questions, 0-68.<br>The higher the better. |

## II. CLINICAL AND RADIOGRAPHIC DETERMINANTS OF TKR OUTCOMES

### *A. Patient-Reported Outcome Measures (PROMs) Scoring Protocols for TKR*

Specific scoring protocols are used to pre- and postoperatively collect PROMs in TKR patients. These protocols are important to TKR outcome evaluation and prediction. They are usually questionnaires completed by patients and assessed by health professionals to evaluate the condition of OA patients. For example, these questionnaires can indicate if a patient's OA status negatively affects their quality of life in terms of both pain and function. Popular scoring protocols are the Knee Society Score (KSS) [23], the Oxford Knee Score (OKS) [24], the Knee Injury and OA Outcome Score (KOOS) [25], and the Western Ontario and McMaster Universities Osteoarthritis Index (WOMAC) [26]. An overview of these protocols is listed in Table I.

KSS [23] includes objective and subjective measures, offering a standardised and systematic approach to assess knee function and pain. This scoring system combines clinical and functional scoring systems. The clinical rating assesses the patient's pain, range of motion, and stability, while the functional scoring system evaluates the patient's capacity to perform daily functions like walking and stair climbing, providing a well-rounded evaluation of postoperative knee performance.

OKS [24] is important for evaluating the functional status and quality of life in individuals undergoing knee surgery, particularly TKR. It focuses on the patient's perspective, capturing their experiences related to pain and function. Comprising 12 questions, OKS assesses various aspects of knee health, including pain severity, stiffness, and limitations in daily activities. The maximum score is 48, indicating the optimal knee function and the absence of knee-related symptoms.

KOOS [25] comprehensively assesses the impact of knee injury and osteoarthritis on an individual's quality of life. It includes five subscales: pain, symptoms, activities of daily living, sport and recreation function, and knee-related quality of life. KOOS does not have a specific overall score. Scores are reported for each of the five subscales instead. This questionnaire provides a detailed and patient-centred evaluation of knee health.

WOMAC [26] measures the three major subscales of osteoarthritis impact: pain, stiffness, and physical function. Comprising 24 items, the WOMAC questionnaire provides a detailed exploration of an individual's experience with osteoarthritis, offering valuable insights into the impact on daily activities and overall quality of life.

### *B. Alignment Biomarkers for TKR Outcomes*

Knee alignment is a key radiographic biomarker for predicting TKR outcomes, influencing surgical planning, implant selection, and postoperative results. In this section, we introduce several knee alignment biomarkers relevant to TKR outcomes. We provide an overview of knee alignment biomarkers and their relations to TKR outcomes in Table II.

Severe preoperative varus deformities have been linked to greater KOOS improvements after TKR [27]. The more severe preoperative OA progressions and higher joint angles are related to better postoperative outcomes measured by WOMAC [28]. However, varus and valgus malalignment have been associated with a higher incidence of revision surgery, both preoperatively [9] and postoperatively [8,29]. For postoperative alignment, neutral limb alignment and higher KSS have been associated only in patients with preoperative non-varus alignment [30]. Postoperative malalignment is also a risk factor for long-term component failure [31]. Not all studies demonstrated clear relationship between alignment and postoperative outcomes. Huijbregts et al. [32] reported that neither mechanical axis nor component alignment is associated with dissatisfaction measured by OKS at one year following TKR.

Tibiofemoral joint deformities could be measured by angles including anatomical femorotibial angle (FTA/TFA) (shown in Fig. 1a), medial proximal tibial angle (MPTA), lateral distal femoral angle (LDFA), and the hip-knee-ankle angle (HKAA)

TABLE II
KNEE ALIGNMENT BIOMARKERS ASSOCIATED WITH TKR OUTCOMES

| Studies | Biomarkers | Conclusions | Number of Participants |
|---|---|---|---|
| [27] | Preoperative varus deformities | More varus deformities lead to higher improvement rates measured by KOOS | 110 patients, 19 males and 91 females |
| [28] | Preoperative OA progression, joint angle | More severe knee OA and higher joint angles were associated with better (lower) postoperative WOMAC scores | 172 patients, 70 males and 102 females |
| [9] | Preoperative varus/valgus alignment | Excessive alignment has a greater risk of failure | 5342 TKRs (3699 patients), 1457 males and 2242 females |
| [8] | Postoperative tibiofemoral alignment | The neutrality of alignment maximize the implant survival | 6070 TKRs (3992 patients), 1556 males and 2436 females |
| [29] | Postoperative coronal alignment | Higher incidence of revision surgery | 6070 TKRs (3992 patients), 1556 males and 2436 females |
| [30] | Postoperative coronal TKR alignment | Neutral limb alignment and higher KSS are associated only in patients with preoperative non-varus alignment | 38 patients, 16 males and 22 females |
| [31] | Postoperative alignment | Postoperative malalignment is a risk factor for long-term component failure | 280 patients, 142 males and 138 females |
| [32] | Postoperative prosthetic alignment | Neither mechanical axis, nor component alignment, is associated with dissatisfaction (OKS) at one year following TKR | 211 patients, 230 TKRs, 105 males and 106 females |

(shown in Fig. 1b), which can also be calculated using their supplementary angles [11-13,33]. Angles are generally measured relative to anatomical or mechanical axes. For lower limbs, the mechanical axis is commonly measured between the centre of the femoral head and the centre of the knee (for the femur) and from the centre of the knee to the ankle (for the tibia), whereas the anatomical axis is a line drawn proximal to distal in the intramedullary canal bisecting the femur and tibia in one-half. Anatomical FTA represents the angle formed by the intersection of the femoral and tibial anatomical axes at the knee, with deviations from the normal angle indicating misalignment or pathology [34]. MPTA measures the angle between the tibial mechanical axis and the proximal tibial joint line, aiding knee deformity evaluation and surgical planning [35]. LDFA is the angle between the femoral mechanical axis and the distal femoral joint line on the lateral side, serving a similar role to the above-mentioned angles [35]. HKAA is defined as the angle between the femoral and tibial mechanical axes [36] and can measure coronal plane knee alignment even better than anatomical FTA, as it includes load distribution within the knee joints [33].

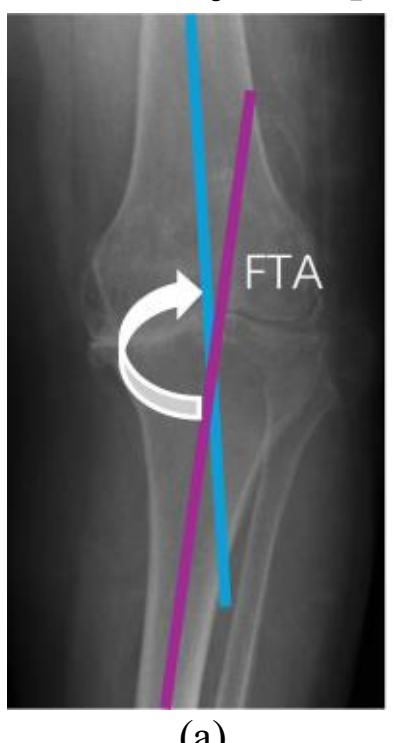

(a)

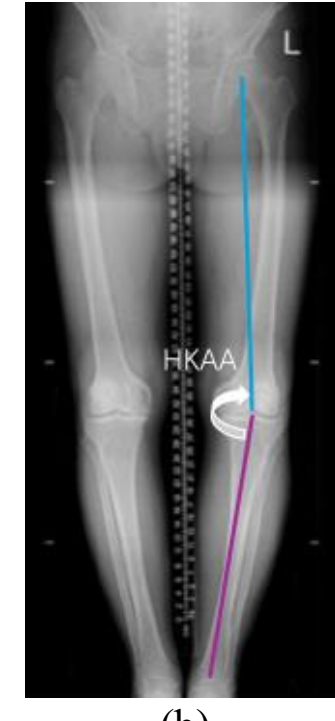

(b)

Fig. 1. An illustration of two widely used angles (FTA and HKAA). In practice, they can also be defined using their supplementary angles [11-13,33].

Patella alignment is another important biomarker for identifying patella-related diseases. Patella height, which refers to the position of the patella relative to the femur and tibia, was identified as another essential radiographic biomarker for TKR outcomes [37]. Patella height can be assessed by measuring the Insall-Salvati index (ISI) [38], the Caton-Deschamps index (CDI) [39], or the Blackburne-Peel index (BPI) [40]. ISI is the ratio of patellar tendon length to patellar length, useful for assessing the patellar position and identifying patellofemoral joint problems. CDI is the ratio of the distance between the anterior corner of the tibial plateau and the most inferior aspect of the patellar articular surface to the patellar articular surface length. It is valuable in assessing patellar instability [41]. BPI is defined as the ratio of the distance between the horizontal line and the inferior aspect of the patellar articular surface to the patellar articular surface length. These above-mentioned ratios, measured in lateral knee radiographs (shown in Fig. 2), are key indicators of patellofemoral disorders such as patella alta and patella baja [37-40].

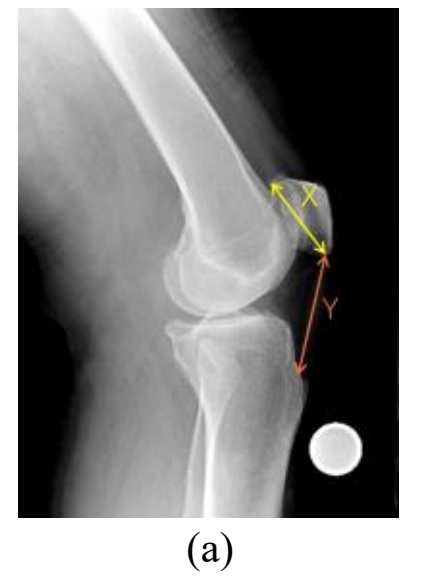

(a)

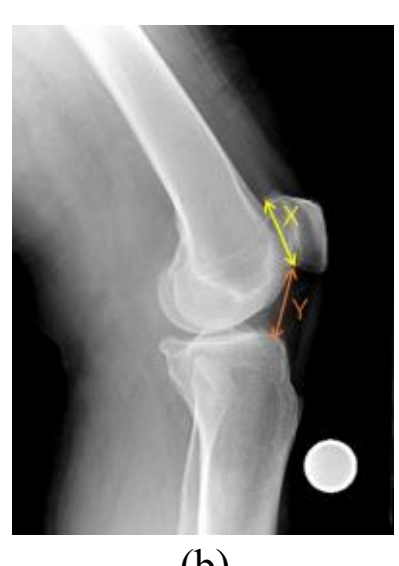

(b)

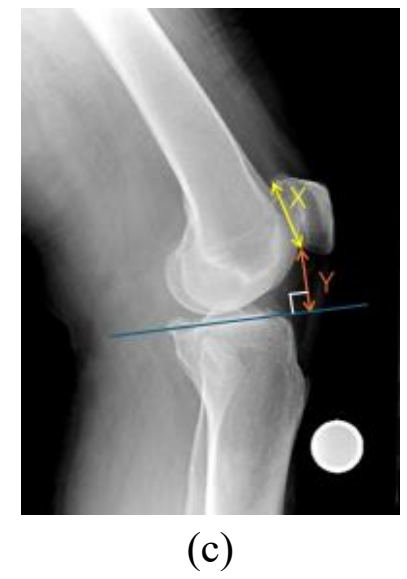

(c)

Fig. 2. An illustration of the three ratios that can be measured from lateral knee radiographs. The ISI (a), CDI (b), and BPI (c) are all calculated by Y/X.

## III. AI-BASED METHODS FOR ASSESSING KNEE ALIGNMENT

Machine learning-based methods such as Random Forests (RF) [42] have been widely used in image processing [43-46].

Deep learning has shown good performance on many problems in various areas of computer vision, such as human face analysis [47,48] and medical imaging applications, including endoscopic dehazing [49-51], image classification [52-54], object segmentation [55-57], disease detection [58-60], digital pathology [61-63], and landmark detection [64-66]. Many popular deep learning models [67] have been used in medical imaging studies, including DenseNet [68], UNet [69], and ResNet [70]. These offer advanced techniques for accurate and efficient processing of radiographs.

### *A. Landmark Detection Solutions*

Landmark detection is a computer vision technique widely applied across imaging tasks. A "landmark" refers to a specific point on an object or within an image that highlights an important feature. Two examples illustrating annotated knee joint landmarks are shown in Fig. 3. Landmark detection is often associated with knee alignment as measurements such as angles can be made based on the detected landmark positions. For example, measuring anatomical FTA or TFA requires first determining the femoral and tibial anatomical axes, which could be defined by landmarks along the femoral condyles, tibial plateau, and shafts. However, manual annotations are subjective and time-consuming. Automated landmark detection can improve the efficiency and the accuracy of knee alignment measurements, thereby supporting more effective diagnostic and treatment planning for knee joint disorders. Often automated landmark detection solutions are generic, and thus, even if not designed specifically for knee radiographs, they can potentially be used for knee landmark detection.

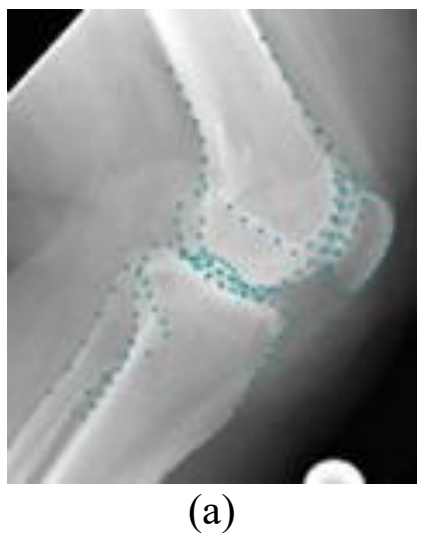
(a)

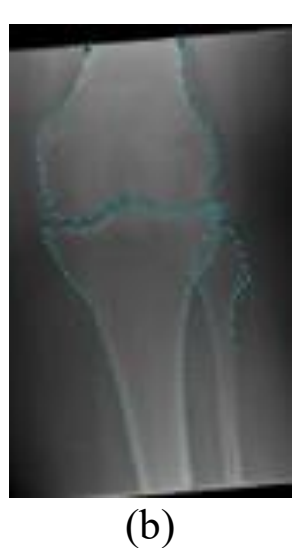
(b)

Fig. 3. Landmark Examples in lateral (a) and AP (b) knee radiographs.

Deformable models can provide effective solutions for automated landmark annotations when combined with model matching techniques [43,44]. These models are flexible curves or surfaces defined within an image domain, offering the necessary degrees of freedom to adapt the model to diverse shapes. Notable examples include Active Shape Models [71] and Active Appearance Models [72], both based on Statistical Shape Models (SSMs). SSMs capture shape variability across a training dataset, enabling the representation and analysis of population-level shape variations. These models generally require training to learn the shape variability or appearance of the target item. Recently, deformable models have been combined with AI techniques like machine learning and deep learning.

Machine learning methods, particularly RF-based approaches, have been widely applied for radiographic landmark detection, often in combination with deformable models. Lindner et al. [44] presented a fully automatic method using random forest regression voting (RFRV) and constrained local models (CLMs) to accurately segment the proximal femur, knee joint, and hand in radiographs. A few candidate positions were produced by a global search with a detector. Each was then refined using an SSM together with local detectors for each model point. Both global and local refinements relied on RF regression to vote for optimal positions, ensuring robustness and accuracy. In two-fold cross-validation on 500 radiographs, 87 knee contour points were localised, achieving a mean point-to-curve error of less than 1 mm in 99% of all cases, demonstrating high accuracy and efficiency.

Deep learning-based methods, particularly neural networks, have also been applied for radiographic landmark detection. Compared with classic landmark regression approaches, heatmap-based landmark regression offers greater robustness. This strategy first generates a probability heatmap for each landmark, and landmark coordinates are subsequently obtained from the heatmap, typically using either the centroid or the maximum response location. Much of the current research has used heatmap-based methods to localise landmarks on pelvis rather than knee radiographs. Davison et al. [73] proposed an automated pelvis landmark detection approach based on heatmap prediction using UNet. They predicted a Gaussian blob for each landmark and estimated displacement of every pixel from each target landmark. Evaluation was performed using relative point-to-point and point-to-curve distances, defined as percentile-based errors normalized to the reference length (femoral shaft width). They reported better performance than [44] with a larger and different dataset including radiographs of hips from children aged between 2 and 11 years. Using 1000 images, their method achieved a mean point-to-curve error of 9.0% at the 99th percentile, outperforming the 17.1% error obtained with the method in [44], though the median point-to-point error was slightly higher (6.7% vs. 5.9%). Huang et al. [74] propose a landmark detection framework for pelvic X-rays that enhances a base detector by topology-aware constraints and pelvis-focused feature enhancement, ultimately refining 14 anatomical landmark coordinates. The detection head predicts heatmaps, which are post-processed through spatial relationship modelling to enforce anatomical structure consistency. Evaluated on two public and one private dataset of around 850 images, the method achieved a best mean radial error of 0.71 mm and a successful detection rate of $\approx$97% at 4 mm, outperforming baseline detectors that lack topology-aware refinement. Pei et al. [75] proposed an attention mechanism of combining multi-dimension information based on separating spatial dimension in the pelvis X-ray landmark detection. The attention modules were inserted into the skipped connections of UNet to form a novel structure. Their method outperformed UNet, HR-Net [76], and CE-Net [77], achieving average point-to-point errors of 3.14 mm versus 3.57 mm, 3.61 mm, and 3.39 mm, respectively. Mulford et al. [78] developed a single CNN model to annotate certain anatomical structures and landmarks on AP pelvis radiographs. A total of 1,100 AP pelvis radiographs were manually annotated by 3 reviewers. The model segmented 22 structures, including 7 landmark points, with Euclidean distance errors for landmarks ranging from 1.9 to 5.6 mm. Of note is that the above mentioned deep learning-based methods did not focus on knee landmark detection. However, similar approaches could be used for knee radiographs to provide biomarkers for knee OA and TKR outcomes.

TABLE III
OVERVIEW OF TECHNICAL SOLUTIONS FOR LANDMARK DETECTION IN LOWER-LIMB RADIOGRAPHS

| Studies | Positions | Techniques | Modalities | Evaluation (Performance) |
|---|---|---|---|---|
| [44] | Proximal femur, knee, hand | RF, SSM | Radiographs | For knee joint: mean point-to-curve distance (99%<1mm) |
| [73] | Proximal femur | UNets | Pelvis radiographs | Relative mean point-to-curve distance (9.0%), Relative median point-to-point distance (6.7%) |
| [74] | Proximal femur | Pelvis Extraction module added to CNNs | Pelvis radiographs | Mean Radial Error: 0.71mm, Successful Detection Rate at 4mm: 97%. |
| [75] | Proximal femur | U-shaped CNNs, attention | Pelvis radiographs | Mean point-to-point distance (3.1350mm) |
| [78] | 22 different structures | UNet | Pelvis radiographs | For the 7-point structures: mean point-to-point distance (1.9mm to 5.6mm, mostly below 3.1mm) |
| [64] | Femur, tibia | Hourglass networks | AP Knee radiographs | Mean point-to-point distance (75%<2mm and 92%<2.5mm for one dataset, 79%<2mm and 93%<2.5mm for another dataset) |
| [80] | Femur, tibia | GCNs | AP Knee radiographs | Mean point-to-point distance (0.84mm, 71.2%<1mm, 93.9%<2mm) |
| [81] | Femur, patella | UNets | Skyline Knee radiographs | Mean point-to-point distance (88.9%<2mm in test set, 82.2%<2mm in hold-out set) |

Currently, few deep learning-based studies have specifically targeted knee X-ray landmark detection. Tiulpin et al. [64] used Hourglass-based networks [79] to regress the knee landmark positions from AP knee radiographs more efficiently. The results demonstrated that over 75% of point-to-point errors were below 2 mm, and over 90% were below 2.5 mm, both of which surpass the performance reported in [44]. Xiao et al. [80] localised anatomical landmarks in AP knee radiographs from Osteoarthritis Initiative (OAI) dataset by combining heatmap regression with a GCN. By representing landmarks as a graph, the model effectively captures structural information, refining landmark coordinates through a cascade of GCNs. Their model achieved a mean point-to-point error of 0.84 mm and the successful detection rate of 71.2% at 1mm. 93.9% of the point-to-point errors were below 2 mm. Fewer studies focused on other radiographic views, such as lateral or skyline. Tuya et al. [81] detected knee landmarks from skyline radiographs and used these landmarks to further measure patellofemoral joint parameters. Their method achieved 88.9% and 82.2% of errors below 2 mm on the original and hold-out test sets, respectively.

A summary of these landmark detection solutions for medical images is shown in Table III.

## B. AI-based Knee Alignment Assessment

### 1) Angles

Many approaches for measuring knee joint angles in full-leg radiographs have been proposed. Moon et al. [82] developed a deep learning-based system to detect lower limb alignment automatically, rapidly, and accurately by using AP standing long-leg radiographic data of lower limbs. The alignment was comprehensively measured by calculating mechanical lateral proximal femoral angle (LPFA), mechanical LDFA, mechanical MPTA, mechanical lateral distal tibial angle (LDTA), mechanical axis deviation, joint line convergence angle (JLCA), mechanical TFA, anatomical medial proximal femoral angle (MPFA), anatomical LDFA, neck shaft angle, anatomical MPTA, anatomical TFA, and the length of femur, tibia, and the full limb. This algorithm includes region of interest (ROI) detection with You Only Look Once (YOLO)v5 [83], segmentation with HarDNet-MSEG [84], and landmark detection. Using radiographs from 770 patients collected between 2016 and 2020, the system achieved results comparable to radiologists, with high agreement metrics (concordance correlation coefficient (CCC), Pearson correlation coefficient (PCC), and intraclass correlation coefficient (ICC) >0.9; mean absolute error (MAE), mean square error (MSE), and root mean square error (RMSE) $\approx$ 1.0). Jo et al. [85] presented a CNN-based anatomical landmark recognition and angle measurement model for femur, tibia, and implant components using full-leg preoperative and postoperative radiographs. 15 anatomical landmarks were marked by two orthopaedic surgeons. Mechanical LDFA, MPTA, JLCA, and HKAA were then measured after extracting the ROIs. The system demonstrated excellent reliability, with inter-observer ICC values $\geqslant$ 0.98 ($p<0.001$) for all angles. Intra-observer ICCs reached 1.00, exceeding those of the orthopaedic specialists (0.97–1.00), indicating that the model matched or surpassed the accuracy of an experienced surgeon with 14 years of practice. Tack et al. [86] proposed a fully automated method for knee alignment quantification using YOLOv4 [87] for ROI detection of the hip, knee, and ankle, followed by ResNet regression of landmark coordinates. Their system yielded HKAA measurements highly consistent with expert annotations (ICC>0.9), supporting its use for automated alignment assessment in full-leg radiographs. Chen et al. [10] used ResNet to automatically measure HKAA on full-leg radiographs without landmark positions. However, their results showed different levels of agreement depending on the imaging method, with knee radiographs taken using a non-positioning frame demonstrating higher reliability (ICC=0.90), compared to those taken within a positioning frame (ICC=0.76).

TABLE IV
OVERVIEW OF TECHNICAL AI-DRIVEN SOLUTIONS FOR KNEE ALIGNMENT ASSESSMENT

| Studies | Measurements | Techniques | Modalities | Landmarks | Evaluation (Performance) |
| --- | --- | --- | --- | --- | --- |
| [82] | LDFA, MPTA, LDTA, LPFA, MPFA, JLCA, TFA | YOLOv5, HarDNET-MSEG | Full-leg X-ray | Required | CCC, PCC, and ICC (<0.9); MAE, MSE, and RMSE (>0.9°) |
| [85] | LDFA, MPTA, JLCA, HKAA | CNNs | Full-leg X-ray | Required | Intra-observer ICC (>0.97); MAE (<0.52°) |
| [86] | HKAA | YOLOv4, ResNet | Full-leg X-ray | Required | Inter-observer ICC (>0.8) |
| [10] | HKAA | ResNet | Full-leg X-ray | Not required | MAE (0.98°), MSE (1.81°), ICC (0.94) in similar test images; MAE (1.56° and 2.10°), MSE (4.10° and 6.63°), ICC (0.76 and 0.90) (for real SynaFlexer™ images and non-positioning frame images, respectively) |
| [33] | Anatomical FTA, HKAA | DenseNet, Inception-ResNet v2 | Knee X-ray | Not required | MAE (0.8° for FTA and 1.7° for HKAA) |
| [11] | Anatomical FTA | RFRV-CLM | Knee X-ray | Required | ICC (0.97/0.78) and MAE (1.2°/1.5°) (pre-/post-operatively). |
| [13] | Anatomical TFA | Hourglass network | Knee X-ray | Required | ICC (0.95/0.86) and MAE (1.4°/1.1°) (pre-/post-operatively). |
| [88] | ISI, CDI | VGG16 | Knee X-ray | Required | ICC (0.91–0.95/0.87–0.96), MAD (0.02–0.05/0.02–0.06), RMSE (0.02–0.07/0.02–0.10) (left/right knee) |
| [89] | CDI, BPI | UNet, YOLO | Knee X-ray | Required | ICC (>0.75), SEM (<0.014) |
| [90] | ISI | ResNet, HR-Net | Knee X-ray | Required | ICC (0.809–0.885) |

However, long-leg radiographs are not always undertaken in clinical practice, and standard AP knee radiographs are often the main imaging modality. Thus, automatic angle prediction from standard knee radiographs is clinically valuable. Wang et al. [33] used different base models to predict anatomical FTA and HKAA angle from posteroanterior (PA) knee radiographs without landmark positions, although HKAA can only be measured with full-limb images. In this study, CNNs with densely connected final layers were trained to analyse PA knee radiographs from the OAI database. Separate models were trained for each angle, with mean squared error as the loss function. Heat maps were used to identify the anatomical features within each image that most contributed to the predicted angles. DenseNet achieved high accuracy, with mean absolute errors of 0.8° for anatomical FTA and 1.7° for HKAA. Agreement was excellent for anatomical FTA (ICC>0.9 across all CNNs), but lower for HKAA (ICC=0.7–0.9), reflecting reduced robustness. Cullen et al. [11] predicted anatomical FTA with two different approaches in AP knee radiographs using anatomical landmark positions detected by RFRV-CLM. Strong agreement was found between the automated and clinical measurements of preoperative anatomical FTA (ICC>0.95). The postoperative agreement between the automated and clinical measurements was lower (ICC=0.7-0.8). Hu et al. [13] applied Hourglass networks to detect anatomical landmark positions in AP knee radiographs and subsequently generate anatomical TFA measurements using the same approaches as [11]. Their system showed higher overall accuracy in landmark detection task compared with RFRV-CLM and stronger agreement between the automated and clinical measurements than RFRV-CLM [11], though the postoperative agreement was also lower than the preoperative one.

*2) Ratios*

AI-based approaches can also be used to measure ratios related to patella alignment, particularly in lateral knee radiographs. Ye et al. [88] developed a CNN-based system for automatic patellar height measurement based on detected landmarks in lateral knee radiographs, focusing on ISI and CDI. Tested on 200 left and 200 right knee radiographs, the system achieved performance comparable to manual measurements by three radiologists (ICC>0.9). Kwolek et al. [89] used YOLO and UNet to detect the ROI and measure patella height with CDI and BPI. Good agreement between the orthopaedic surgeons' measurements and results of their algorithm was achieved (ICC>0.75, standard error for single measurement (SEM) <0.014). Liu et al. [90] used ResNet and HR-Net to automatically measure ISI. The system performed excellent in keypoint detection tasks and was highly consistent with the manual measurements of ISI (ICC, 0.809–0.885).

A summary of AI-based methods for measuring knee alignment angles and ratios is shown in Table IV.

### *C. Commercial Products for Automated Knee Alignment Assessment*

Many companies have developed AI-based solutions for knee image analysis and primarily focus on fracture detection. For instance, Mediaire[1] developed a software system for AI-based evaluation of knee MRIs through automatic detection and classification of cartilage damage according to the International Cartilage Regeneration & Joint Preservation Society (ICRS) grading system. RBfracture developed by Radiobotics[2] can detect fractures across the appendicular skeleton, ribs and pelvis and is integrated in a standard reading environment. Aztrauma

[1]https://mediaire.ai/en/mdknee/

[2]https://radiobotics.com/solutions/rbfracture/

TABLE V
AVAILABLE SOFTWARE PRODUCTS FOR RADIOGRAPHIC KNEE ALIGNMENT ASSESSMENT

| Companies | Measurements | References | Modalities | Evaluation (Performance) |
|---|---|---|---|---|
| ImageBiopsy Lab | HKAA, TFA, JLCA, LDFA, MPTA, LDTA, LPFA | [91], [92], [93] | Full-leg X-ray | ICC (>0.9); MAE(<1°) |
| Gleamer | HKAA | [94] | Full-leg X-ray | ICC (>0.99); RMSE (0.37); MSE (0.30) |
| Milvue | HKAA | N/A | Full-leg X-ray | N/A |

developed by Azmed[3] can be applied for the same purpose.

Several automated solutions specifically address knee alignment assessment and have been integrated into picture archiving and communication systems (PACS). However, these products typically rely on full-leg radiographs to measure angles such as HKAA. ImageBiopsy Lab[4] has developed an AI-powered software solution to automate and standardise musculoskeletal imaging diagnostics. Their software automatically measures JLCA, TFA, and HKAA, achieving near-perfect agreement with reference standards (ICC close to 1; MAEs $\leqslant 1^\circ$) [91,92]. Moreover, the intra-observer ICC for automated approach is better than manual ones [93], demonstrating more reliable measurements. Gleamer[5] has developed AI-driven software like BoneMetrics, which automates measurements including knee alignment on full-leg standing radiographs, delivering accurate and reproducible HKAA values with excellent agreement compared to ground truth [94]. TechCare Bones developed by Milvue[6] also provides an automated solution to assessing knee alignment including HKAA from full-leg radiographs, but no literature is available to provide evidence of its performance.

For the analysis of standard knee radiographs, current commercial products are typically limited to grading knee osteoarthritis. Although full-leg radiographs provide more comprehensive anatomical information, it is clinically valuable to explore the feasibility of alignment measurement using standard knee radiographs.

Currently available commercial products for radiographic knee alignment measurement are summarised in Table V.

## IV. DISCUSSIONS AND CONCLUSIONS

Knee alignment is highly associated with TKR outcomes. While varus and valgus deformities may lead to greater postoperative improvement, they can also be linked to a higher incidence of revision surgery. Additionally, postoperative malalignment is a risk factor for long-term component failure. However, no clear relationship was found between knee alignment and postoperative outcomes when assessed using OKS. It would be valuable to provide further evidence on the relationship between alignment and postoperative outcomes using different scoring protocols to enhance the robustness and clinical relevance of the findings.

Tibiofemoral alignment is typically measured using angles such as FTA, TFA, MPTA, LDFA, and HKAA, while patella height is evaluated through ratios including ISI, CDI, and BPI. Landmark detection is commonly used to identify key points in radiographs, enabling subsequent knee alignment measurements. Both machine learning and deep learning-based approaches, such as RFRV-CLM and CNNs, have been developed for this purpose. CNNs may offer greater reliability and robustness in landmark detection but are not yet more accurate than RFRV-CLM. Currently, a more accurate pipeline should involve RFRV-CLM to refine the point positions initialised by CNNs.

AI-based techniques could accurately and efficiently help with the knee alignment assessments. Many approaches showed excellent accuracy comparable to the manual measurements. When measuring some of the angles, such as HKAA, full-leg radiographs are traditionally required. However, acquiring and pre-processing these images can be time-consuming, and they result in increased radiation exposure. Therefore, it would be clinically valuable to develop innovative approaches for predicting these measurements using standard knee radiographs. Current approaches for HKAA prediction from standard knee radiographs remain unstable, highlighting the need for more robust and reliable models. Most existing alignment assessments methods rely on landmark detection, though direct angle prediction could provide an alternative pathway. Furthermore, many studies have not included radiologist-provided manual annotations or tested their models on independent datasets, limiting confidence in generalisability. Future work should emphasise external validation on unseen imaging data to better demonstrate clinical applicability and ensure reliable performance in diverse practice settings.

Most current methods for knee alignment assessment focus on the AP and lateral views, with only a few studies addressing skyline images [95] for parameters such as patellar tilt. Unlike 3D modalities such as MRI, combining different X-ray views is more challenging, and most existing approaches for landmark detection and alignment measurement only analyse a single view. Developing methods to combine multiple radiographic views could provide more comprehensive information [96].

While the link between knee alignment and TKR outcomes is well established, no automated approaches currently predict TKR outcomes directly from radiographic alignment. Developing end-to-end systems that either integrate automated alignment measurement with outcome prediction, or directly infer outcomes from radiographs, represents a promising research direction.

Many AI-based knee alignment assessment solutions have been proposed, however, only full-leg-based knee alignment assessment has been integrated with PACS in clinical practice.

[3] https://www.azmed.co/azproducts-pages/aztrauma
[4] https://www.imagebiopsy.com/
[5] https://www.gleamer.ai/
[6] https://www.milvue.com/en/solutions/techcarebones/

Seamlessly incorporating the proposed AI-based knee alignment assessment tools for knee radiographs into existing clinical systems is crucial, as these are commonly used in clinical practice and expose patients to less radiation.

## References

[1] Jin, Z., Wang, D., Zhang, H., Liang, J., Feng, X., Zhao, J. and Sun, L., 2020. Incidence trend of five common musculoskeletal disorders from 1990 to 2017 at the global, regional and national level: results from the global burden of disease study 2017. Annals of the rheumatic diseases, 79(8), pp.1014-1022.

[2] Litwic, A., Edwards, M.H., Dennison, E.M. and Cooper, C., 2013. Epidemiology and burden of osteoarthritis. British medical bulletin, 105(1), pp.185-199.

[3] Langworthy, M., Dasa, V. and Spitzer, A.I., 2024. Knee osteoarthritis: disease burden, available treatments, and emerging options. Therapeutic Advances in Musculoskeletal Disease, 16, p.1759720X241273009.

[4] Jin, X., Liang, W., Zhang, L., Cao, S., Yang, L. and Xie, F., 2023. Economic and humanistic burden of osteoarthritis: an updated systematic review of large sample studies. Pharmacoeconomics, 41(11), pp.1453-1467.

[5] Özden, V.E., Osman, W.S., Morii, T., Pastor, J.C.M., Abdelaal, A.M. and Younis, A.S., 2025. What percentage of patients are dissatisfied post-primary total hip and total knee arthroplasty?. The Journal of Arthroplasty, 40(2), pp.S55-S56.

[6] DeFrance, M.J. and Scuderi, G.R., 2023. Are 20% of patients actually dissatisfied following total knee arthroplasty? A systematic review of the literature. The Journal of arthroplasty, 38(3), pp.594-599.

[7] Meadows, K.A., 2011. Patient-reported outcome measures: an overview. British journal of community nursing, 16(3), pp.146-151.

[8] Ritter, M.A., Davis, K.E., Meding, J.B., Pierson, J.L., Berend, M.E. and Malinzak, R.A., 2011. The effect of alignment and BMI on failure of total knee replacement. JBJS, 93(17), pp.1588-1596.

[9] Ritter, M.A., Davis, K.E., Davis, P., Farris, A., Malinzak, R.A., Berend, M.E. and Meding, J.B., 2013. Preoperative malalignment increases risk of failure after total knee arthroplasty. JBJS, 95(2), pp.126-131.

[10] Chen, K., Stotter, C., Lepenik, C., Klestil, T., Salzlechner, C. and Nehrer, S., 2025. Frontal plane mechanical leg alignment estimation from knee x-rays using deep learning. Osteoarthritis and Cartilage Open, 7(1), p.100551.

[11] Cullen, D., Thompson, P., Johnson, D. and Lindner, C., 2025. An AI-based system for fully automated knee alignment assessment in standard AP knee radiographs. The Knee, 54, pp.99-110.

[12] Hu, Z., Cullen, D., Thompson, P., Johnson, D., Tiulpin, A., Cootes, T.F. and Lindner, C., 2025. Automated measurements of knee alignment with deep learning: accuracy and reliability. Osteoarthritis and Cartilage, 33, pp.S100-S101.

[13] Hu, Z., Cullen, D., Thompson, P., Johnson, D., Bian, C., Tiulpin, A., Cootes, T. and Lindner, C., 2025, September. Deep learning-based alignment measurement in knee radiographs. In International Conference on Medical Image Computing and Computer-Assisted Intervention (pp. 121-130). Cham: Springer Nature Switzerland.

[14] Kokkotis, C., Moustakidis, S., Papageorgiou, E., Giakas, G. and Tsaopoulos, D.E., 2020. Machine learning in knee osteoarthritis: A review. Osteoarthritis and Cartilage Open, 2(3), p.100069.

[15] Yeoh, P.S.Q., Lai, K.W., Goh, S.L., Hasikin, K., Hum, Y.C., Tee, Y.K. and Dhanalakshmi, S., 2021. Emergence of deep learning in knee osteoarthritis diagnosis. Computational intelligence and neuroscience, 2021(1), p.4931437.

[16] Binvignat, M., Pedoia, V., Butte, A.J., Louati, K., Klatzmann, D., Berenbaum, F., Mariotti-Ferrandiz, E. and Sellam, J., 2022. Use of machine learning in osteoarthritis research: a systematic literature review. RMD open, 8(1), p.e001998.

[17] Ramazanian, T., Fu, S., Sohn, S., Taunton, M.J. and Kremers, H.M., 2023. Prediction models for knee osteoarthritis: review of current models and future directions. Archives of Bone and Joint Surgery, 11(1), p.1.

[18] Jamshidi, A., Pelletier, J.P. and Martel-Pelletier, J., 2019. Machine-learning-based patient-specific prediction models for knee osteoarthritis. Nature Reviews Rheumatology, 15(1), pp.49-60.

[19] Gan, H.S., Ramlee, M.H., Wahab, A.A., Lee, Y.S. and Shimizu, A., 2021. From classical to deep learning: review on cartilage and bone segmentation techniques in knee osteoarthritis research. Artificial Intelligence Review, 54(4), pp.2445-2494.

[20] Ebrahimkhani, S., Jaward, M.H., Cicuttini, F.M., Dharmaratne, A., Wang, Y. and de Herrera, A.G.S., 2020. A review on segmentation of knee articular cartilage: from conventional methods towards deep learning. Artificial intelligence in medicine, 106, p.101851.

[21] Ahmed, S.M. and Mstafa, R.J., 2022. A comprehensive survey on bone segmentation techniques in knee osteoarthritis research: From conventional methods to deep learning. Diagnostics, 12(3), p.611.

[22] Ridhma, Kaur, M., Sofat, S. and Chouhan, D.K., 2021. Review of automated segmentation approaches for knee images. IET Image Processing, 15(2), pp.302-324.

[23] Caplan, N. and Kader, D.F., 2013. Rationale of the Knee Society clinical rating system. In Classic Papers in Orthopaedics (pp. 197-199). London: Springer London.

[24] Dawson, J., Fitzpatrick, R., Murray, D. and Carr, A., 1998. Questionnaire on the perceptions of patients about total knee replacement. The Journal of Bone & Joint Surgery British Volume, 80(1), pp.63-69.

[25] Roos, E.M. and Lohmander, L.S., 2003. The Knee injury and Osteoarthritis Outcome Score (KOOS): from joint injury to osteoarthritis. Health and quality of life outcomes, 1(1), p.64.

[26] Bellamy, N., Buchanan, W.W., Goldsmith, C.H., Campbell, J. and Stitt, L.W., 1988. Validation study of WOMAC: a health status instrument for measuring clinically important patient relevant outcomes to antirheumatic drug therapy in patients with osteoarthritis of the hip or knee. The Journal of rheumatology, 15(12), pp.1833-1840.

[27] Toguchi, K., Nakajima, A., Akatsu, Y., Sonobe, M., Yamada, M., Takahashi, H., Saito, J., Aoki, Y., Suguro, T. and Nakagawa, K., 2020. Predicting clinical outcomes after total knee arthroplasty from preoperative radiographic factors of the knee osteoarthritis. BMC Musculoskeletal Disorders, 21(1), p.9.

[28] Kahn, T.L., Soheili, A. and Schwarzkopf, R., 2013. Outcomes of total knee arthroplasty in relation to preoperative patient-reported and radiographic measures: data from the osteoarthritis initiative. Geriatric orthopaedic surgery & rehabilitation, 4(4), pp.117-126.

[29] Fang, D.M., Ritter, M.A. and Davis, K.E., 2009. Coronal alignment in total knee arthroplasty: just how important is it?. The Journal of arthroplasty, 24(6), pp.39-43.

[30] Slevin, O., Hirschmann, A., Schiapparelli, F.F., Amsler, F., Huegli, R.W. and Hirschmann, M.T., 2018. Neutral alignment leads to higher knee society scores after total knee arthroplasty in preoperatively non-varus patients: a prospective clinical study using 3D-CT. Knee Surgery, Sports Traumatology, Arthroscopy, 26(6), pp.1602-1609.

[31] Parratte, S., Pagnano, M.W., Trousdale, R.T. and Berry, D.J., 2010. Effect of postoperative mechanical axis alignment on the fifteen-year survival of modern, cemented total knee replacements. JBJS, 92(12), pp.2143-2149.

[32] Huijbregts, H.J., Khan, R.J., Fick, D.P., Jarrett, O.M. and Haebich, S., 2016. Prosthetic alignment after total knee replacement is not associated with dissatisfaction or change in Oxford Knee Score: a multivariable regression analysis. The Knee, 23(3), pp.535-539.

[33] Wang, J., Hall, T.A., Musbahi, O., Jones, G.G. and van Arkel, R.J., 2023. Predicting hip-knee-ankle and femorotibial angles from knee radiographs with deep learning. The Knee, 42, pp.281-288.

[34] McDaniel, G., Mitchell, K.L., Charles, C. and Kraus, V.B., 2010. A comparison of five approaches to measurement of anatomic knee alignment from radiographs. Osteoarthritis and cartilage, 18(2), pp.273-277.

[35] Lee, S.H., Yoo, J.H., Kwak, D.K., Kim, S.H., Chae, S.K. and Moon, H.S., 2024. The posterior tibial slope affects the measurement reliability regarding the radiographic parameter of the knee. BMC Musculoskeletal Disorders, 25(1), p.202.

[36] Paley, D., 2014. Principles of deformity correction. Springer.

[37] Gaillard, R., Bankhead, C., Budhiparama, N., Batailler, C., Servien, E. and Lustig, S., 2019. Influence of patella height on total knee arthroplasty: outcomes and survival. The Journal of arthroplasty, 34(3), pp.469-477.

[38] Insall, J. and Salvati, E., 1971. Patella position in the normal knee joint. Radiology, 101(1), pp.101-104.

[39] Caton, J., Deschamps, G., Chambat, P., Lerat, J.L. and Dejour, H., 1982. Patella infera. Apropos of 128 cases. Revue de chirurgie orthopedique et reparatrice de l'appareil moteur, 68(5), pp.317-325.

[40] Blackburne, J.S. and Peel, T.E., 1977. A new method of measuring patellar height. The Journal of Bone & Joint Surgery British Volume, 59(2), pp.241-242.

[41] Paul, R.W., Brutico, J.M., Wright, M.L., Erickson, B.J., Tjoumakaris, F.P., Freedman, K.B. and Bishop, M.E., 2021. Strong agreement between

magnetic resonance imaging and radiographs for Caton–Deschamps index in patients with patellofemoral instability. Arthroscopy, Sports Medicine, and Rehabilitation, 3(6), pp.e1621-e1628.

[42] Breiman, L., 2001. Random forests. Machine learning, 45(1), pp.5-32.

[43] Lindner, C., Bromiley, P.A., Ionita, M.C. and Cootes, T.F., 2014. Robust and accurate shape model matching using random forest regression-voting. IEEE transactions on pattern analysis and machine intelligence, 37(9), pp.1862-1874.

[44] Lindner, C., Thiagarajah, S., Wilkinson, J.M., arcOGEN Consortium, Wallis, G.A. and Cootes, T.F., 2013, September. Accurate bone segmentation in 2D radiographs using fully automatic shape model matching based on regression-voting. In International conference on medical image computing and computer-assisted intervention (pp. 181-189). Berlin, Heidelberg: Springer Berlin Heidelberg.

[45] Cootes, T.F., Ionita, M.C., Lindner, C. and Sauer, P., 2012, October. Robust and accurate shape model fitting using random forest regression voting. In European conference on computer vision (pp. 278-291). Berlin, Heidelberg: Springer Berlin Heidelberg.

[46] Lindner, C., Thiagarajah, S., Wilkinson, J.M., Wallis, G.A., Cootes, T.F. and arcOGEN Consortium, 2013. Fully automatic segmentation of the proximal femur using random forest regression voting. IEEE transactions on medical imaging, 32(8), pp.1462-1472.

[47] Wu, W., Qian, C., Yang, S., Wang, Q., Cai, Y. and Zhou, Q., 2018. Look at boundary: A boundary-aware face alignment algorithm. In Proceedings of the IEEE conference on computer vision and pattern recognition (pp. 2129-2138).

[48] Dou, H., Chen, C., Hu, X., Xuan, Z., Hu, Z. and Peng, S., 2020, October. PCA-SRGAN: Incremental orthogonal projection discrimination for face super-resolution. In Proceedings of the 28th ACM international conference on multimedia (pp. 1891-1899).

[49] Zhou, Y., Hu, Z., Xuan, Z., Wang, Y. and Hu, X., 2022. Synchronizing detection and removal of smoke in endoscopic images with cyclic consistency adversarial nets. IEEE/ACM Transactions on Computational Biology and Bioinformatics, 21(4), pp.670-680.

[50] Hu, Z. and Hu, X., 2021, November. Cycle-consistent adversarial networks for smoke detection and removal in endoscopic images. In 2021 43rd Annual International Conference of the IEEE Engineering in Medicine & Biology Society (EMBC) (pp. 3070-3073). IEEE.

[51] Chen, L., Tang, W., John, N.W., Wan, T.R. and Zhang, J.J., 2019. De-smokeGCN: generative cooperative networks for joint surgical smoke detection and removal. IEEE transactions on medical imaging, 39(5), pp.1615-1625.

[52] Yue, Y. and Li, Z., 2024. Medmamba: Vision mamba for medical image classification. arXiv preprint arXiv:2403.03849.

[53] Zheng, F., Cao, J., Yu, W., Chen, Z., Xiao, N. and Lu, Y., 2024. Exploring low-resource medical image classification with weakly supervised prompt learning. Pattern Recognition, 149, p.110250.

[54] Zhang, J., Xie, Y., Wu, Q. and Xia, Y., 2019. Medical image classification using synergic deep learning. Medical image analysis, 54, pp.10-19.

[55] Chen, J., Mei, J., Li, X., Lu, Y., Yu, Q., Wei, Q., Luo, X., Xie, Y., Adeli, E., Wang, Y. and Lungren, M.P., 2024. TransUNet: Rethinking the U-Net architecture design for medical image segmentation through the lens of transformers. Medical Image Analysis, 97, p.103280.

[56] Chen, C., Miao, J., Wu, D., Zhong, A., Yan, Z., Kim, S., Hu, J., Liu, Z., Sun, L., Li, X. and Liu, T., 2024. Ma-sam: Modality-agnostic sam adaptation for 3d medical image segmentation. Medical Image Analysis, 98, p.103310.

[57] Chen, L., Bentley, P., Mori, K., Misawa, K., Fujiwara, M. and Rueckert, D., 2018. DRINet for medical image segmentation. IEEE transactions on medical imaging, 37(11), pp.2453-2462.

[58] Kar, N.K., Jana, S., Rahman, A., Ashokrao, P.R. and Mangai, R.A., 2024, July. Automated intracranial hemorrhage detection using deep learning in medical image analysis. In 2024 International Conference on Data Science and Network Security (ICDSNS) (pp. 1-6). IEEE.

[59] Javed, R., Abbas, T., Khan, A.H., Daud, A., Bukhari, A. and Alharbey, R., 2024. Deep learning for lungs cancer detection: a review. Artificial Intelligence Review, 57(8), p.197.

[60] Hirvasniemi, J., Runhaar, J., van der Heijden, R.A., Zokaeinikoo, M., Yang, M., Li, X., Tan, J., Rajamohan, H.R., Zhou, Y., Deniz, C.M. and Caliva, F., 2023. The KNee OsteoArthritis Prediction (KNOAP2020) challenge: An image analysis challenge to predict incident symptomatic radiographic knee osteoarthritis from MRI and X-ray images. Osteoarthritis and cartilage, 31(1), pp.115-125.

[61] Bian, C., Wang, Y., Lu, Z., An, Y., Wang, H., Kong, L., Du, Y. and Tian, J., 2021. ImmunoAIzer: A deep learning-based computational framework to characterize cell distribution and gene mutation in tumor microenvironment. Cancers, 13(7), p.1659.

[62] Bian, C., Wang, Y., An, Y., Wang, H., Du, Y. and Tian, J., 2021, February. A computational prediction method based on modified U-Net for cell distribution in tumor microenvironment. In Medical Imaging 2021: Digital Pathology (Vol. 11603, pp. 64-69). SPIE.

[63] Mandair, D., Reis-Filho, J.S. and Ashworth, A., 2023. Biological insights and novel biomarker discovery through deep learning approaches in breast cancer histopathology. NPJ breast cancer, 9(1), p.21.

[64] Tiulpin, A., Melekhov, I. and Saarakkala, S., 2019. KNEEL: Knee anatomical landmark localization using hourglass networks. In Proceedings of the IEEE/CVF International Conference on Computer Vision Workshops (pp. 0-0).

[65] Alansary, A., Oktay, O., Li, Y., Le Folgoc, L., Hou, B., Vaillant, G., Kamnitsas, K., Vlontzos, A., Glocker, B., Kainz, B. and Rueckert, D., 2019. Evaluating reinforcement learning agents for anatomical landmark detection. Medical image analysis, 53, pp.156-164.

[66] Wang, Y., Wu, W., Christelle, M., Sun, M., Wen, Z., Lin, Y., Zhang, H. and Xu, J., 2024. Automated localization of mandibular landmarks in the construction of mandibular median sagittal plane. European Journal of Medical Research, 29(1), p.84.

[67] LeCun, Y., Bengio, Y. and Hinton, G., 2015. Deep learning. nature, 521(7553), pp.436-444.

[68] Iandola, F., Moskewicz, M., Karayev, S., Girshick, R., Darrell, T. and Keutzer, K., 2014. Densenet: Implementing efficient convnet descriptor pyramids. arXiv preprint arXiv:1404.1869.

[69] Ronneberger, O., Fischer, P. and Brox, T., 2015, October. U-net: Convolutional networks for biomedical image segmentation. In International Conference on Medical image computing and computer-assisted intervention (pp. 234-241). Cham: Springer international publishing.

[70] He, K., Zhang, X., Ren, S. and Sun, J., 2016. Deep residual learning for image recognition. In Proceedings of the IEEE conference on computer vision and pattern recognition (pp. 770-778).

[71] Cootes, T.F., Taylor, C.J., Cooper, D.H. and Graham, J., 1995. Active shape models-their training and application. Computer vision and image understanding, 61(1), pp.38-59.

[72] Cootes, T.F., Edwards, G.J. and Taylor, C.J., 2002. Active appearance models. IEEE Transactions on pattern analysis and machine intelligence, 23(6), pp.681-685.

[73] Davison, A.K., Lindner, C., Perry, D.C., Luo, W., Medical Student Annotation Collaborative and Cootes, T.F., 2018, September. Landmark localisation in radiographs using weighted heatmap displacement voting. In International workshop on computational methods and clinical applications in musculoskeletal imaging (pp. 73-85). Cham: Springer International Publishing.

[74] Huang, Z., Li, H., Shao, S., Zhu, H., Hu, H., Cheng, Z., Wang, J. and Kevin Zhou, S., 2024. Pele scores: pelvic x-ray landmark detection with pelvis extraction and enhancement. International Journal of Computer Assisted Radiology and Surgery, 19(5), pp.939-950.

[75] Pei, Y., Mu, L., Xu, C., Li, Q., Sen, G., Sun, B., Li, X. and Li, X., 2023. Learning-based landmark detection in pelvis x-rays with attention mechanism: data from the osteoarthritis initiative. Biomedical Physics & Engineering Express, 9(2), p.025001.

[76] Wang, J., Sun, K., Cheng, T., Jiang, B., Deng, C., Zhao, Y., Liu, D., Mu, Y., Tan, M., Wang, X. and Liu, W., 2020. Deep high-resolution representation learning for visual recognition. IEEE transactions on pattern analysis and machine intelligence, 43(10), pp.3349-3364.

[77] Gu, Z., Cheng, J., Fu, H., Zhou, K., Hao, H., Zhao, Y., Zhang, T., Gao, S. and Liu, J., 2019. Ce-net: Context encoder network for 2d medical image segmentation. IEEE transactions on medical imaging, 38(10), pp.2281-2292.

[78] Mulford, K.L., Johnson, Q.J., Mujahed, T., Khosravi, B., Rouzrokh, P., Mickley, J.P., Taunton, M.J. and Wyles, C.C., 2023. A deep learning tool for automated landmark annotation on hip and pelvis radiographs. The Journal of Arthroplasty, 38(10), pp.2024-2031.

[79] Newell, A., Yang, K. and Deng, J., 2016, September. Stacked hourglass networks for human pose estimation. In European conference on computer vision (pp. 483-499). Cham: Springer International Publishing.

[80] Xiao, J., Dang, K. and Ding, X., 2023, October. Anatomical landmark localization for knee x-ray images via heatmap regression refined with graph convolutional network. In 2023 16th International Congress on Image and Signal Processing, BioMedical Engineering and Informatics (CISP-BMEI) (pp. 1-6). IEEE.

[81] Tuya, E., Nai, R., Liu, X., Wang, C., Liu, J., Li, S., Huang, J., Yu, J., Zhang, Y., Liu, W. and Zhang, X., 2023. Automatic measurement of the patellofemoral joint parameters in the Laurin view: a deep learning–based approach. European Radiology, 33(1), pp.566-577.

[82] Moon, K.R., Lee, B.D. and Lee, M.S., 2023. A deep learning approach for fully automated measurements of lower extremity alignment in radiographic images. Scientific Reports, 13(1), p.14692.

[83] Jocher, G., Chaurasia, A., Stoken, A., Borovec, J., Kwon, Y., Michael, K., Fang, J., Yifu, Z., Wong, C., Montes, D. and Wang, Z., 2022. ultralytics/yolov5: v7. 0-yolov5 sota realtime instance segmentation. Zenodo.

[84] Huang, C.H., Wu, H.Y. and Lin, Y.L., 2021. Hardnet-mseg: A simple encoder-decoder polyp segmentation neural network that achieves over 0.9 mean dice and 86 fps. arXiv preprint arXiv:2101.07172.

[85] Jo, C., Hwang, D., Ko, S., Yang, M.H., Lee, M.C., Han, H.S. and Ro, D.H., 2023. Deep learning-based landmark recognition and angle measurement of full-leg plain radiographs can be adopted to assess lower extremity alignment. Knee Surgery, Sports Traumatology, Arthroscopy, 31(4), pp.1388-1397.

[86] Tack, A., Preim, B. and Zachow, S., 2021. Fully automated assessment of knee alignment from full-leg X-rays employing a "YOLOv4 And Resnet Landmark regression Algorithm"(YARLA): data from the Osteoarthritis Initiative. Computer Methods and Programs in Biomedicine, 205, p.106080.

[87] Bochkovskiy, A., Wang, C.Y. and Liao, H.Y.M., 2020. Yolov4: Optimal speed and accuracy of object detection. arXiv preprint arXiv:2004.10934.

[88] Ye, Q., Shen, Q., Yang, W., Huang, S., Jiang, Z., He, L. and Gong, X., 2020. Development of automatic measurement for patellar height based on deep learning and knee radiographs. European Radiology, 30(9), pp.4974-4984.

[89] Kwolek, K., Grzelecki, D., Kwolek, K., Marczak, D., Kowalczewski, J. and Tyrakowski, M., 2023. Automated patellar height assessment on high-resolution radiographs with a novel deep learning-based approach. World Journal of Orthopedics, 14(6), p.387.

[90] Liu, Z., Wu, J., Gao, X., Qin, Z., Tian, R. and Wang, C., 2024. Deep learning-based automatic measurement system for patellar height: a multicenter retrospective study. Journal of Orthopaedic Surgery and Research, 19(1), p.324.

[91] Archer, H., Reine, S., Xia, S., Vazquez, L.C., Ashikyan, O., Pezeshk, P., Kohli, A., Xi, Y., Wells, J.E., Hummer, A. and Difranco, M., 2024. Deep learning generated lower extremity radiographic measurements are adequate for quick assessment of knee angular alignment and leg length determination. Skeletal radiology, 53(5), pp.923-933.

[92] Schwarz, G.M., Simon, S., Mitterer, J.A., Frank, B.J., Aichmair, A., Dominkus, M. and Hofstaetter, J.G., 2022. Artificial intelligence enables reliable and standardized measurements of implant alignment in long leg radiographs with total knee arthroplasties. Knee Surgery, Sports Traumatology, Arthroscopy, 30(8), pp.2538-2547.

[93] Stotter, C., Klestil, T., Chen, K., Hummer, A., Salzlechner, C., Angele, P. and Nehrer, S., 2023. Artificial intelligence-based analyses of varus leg alignment and after high tibial osteotomy show high accuracy and reproducibility. Knee Surgery, Sports Traumatology, Arthroscopy, 31(12), pp.5885-5895.

[94] Lassalle, L., Regnard, N.E., Durteste, M., Ventre, J., Marty, V., Clovis, L., Zhang, Z., Nitche, N., Ducarouge, A., Laredo, J.D. and Guermazi, A., 2024. Evaluation of a deep learning software for automated measurements on full-leg standing radiographs. Knee Surgery & Related Research, 36(1), p.40.

[95] Hunter, S., Omiye, T. and Lee, M., 2022. Automatic Measurement of Patellar Tilt using Deep Learning Methods.

[96] Minciullo, L., Thomson, J. and Cootes, T.F., 2017, March. Combination of lateral and PA view radiographs to study development of knee OA and associated pain. In Medical Imaging 2017: Computer-Aided Diagnosis (Vol. 10134, pp. 255-261). SPIE.